# Thermodynamic and structural consensus principle predicts mature miRNA location and structure, categorizes conserved interspecies miRNA subgroups, and hints new possible mechanisms of miRNA maturization

Molei Tao (1)

((1) Control and Dynamical Systems, California Institute of Technology, Pasadena, CA 91125, USA)
mtao@caltech.edu

## Abstract

Although conservation of thermodynamics is much less studied than of sequences and structures, thermodynamic details are biophysical features different from but as important as structural features. As a succession of previous research which revealed the important relationships between thermodynamic features and miRNA maturization, this article applies interspecies conservation of miRNA thermodynamics and structures to study miRNA maturization.

According to a thermodynamic and structural consensus principle, miRBase is categorized by conservation subgroups, which imply various functions. These subgroups are divided without the introduction of functional information. This suggests the consistency between the two processes of miRNA maturization and functioning.

Different from prevailing methods which predict extended miRNA precursors, a learning-based algorithm is proposed to predict ~22bp mature parts of 2780 test miRNA genes in 44 species with a rate of 79.4%. This is the first attempt of a general interspecies prediction of mature miRNAs.

Suboptimal structures that most fit the consensus thermodynamic and structural profiles are chosen to improve structure prediction.

Distribution of miRNA locations on corresponding pri-miRNA stem-loop structures is then studied. Existing research on Drosha cleavage site is not generally true across species. Instead, the distance between mature miRNA and center loop normalized by stem length is a more conserved structural feature in animals, and the normalized distance between mature miRNA and ss-RNA tail is the counterpart in plants. This suggests two possibly-updating mechanisms of miRNA maturization in animals and plants.

All in all, conservations of thermodynamics together with other features are shown closely related to miRNA maturization.

## Introduction

MicroRNAs (miRNAs) are endogenous RNAs that play important regulatory roles by targeting mRNAs for cleavage or translational repression. Being non-coding genes, miRNAs is a big class of gene regulatory molecules which influence the output of many protein-coding genes [Bartel, 2004]. It is widely acknowledged that miRNA exists in multicellular organisms, including mammals, fish, worms, flies, cress, rice, etc. [Pasquinelli et al., 2000; Lagos-Quintana et al., 2001, 2002, 2003; Mourelatos et al., 2002; Ambros et al., 2003; Aravin et al., 2003; Dostie et al., 2003; Houbaviy et al., 2003; Kim et al., 2004; Lim et al., 2003a, 2003b; Michael et al., 2003; Park et al., 2002; Reinhart et al., 2002; Palatnik et al., 2003]. Although its existence in prokaryotes is still under investigation and actually depends on the definition, miRNAs is a very general regulatory mechanism. For instance, miRNAs are also found in Epstein-Barr virus [Pfeffer et al., 2004].

The general existence of miRNA might be due to the importance of the regulatory roles that it plays. For example, it is estimated that human might only have 25000 genes [International Human Genome Sequencing Consortium, 2004], and its full complexity is realized with the aid of non-coding RNAs [Claverie, 2005], of which miRNA is one major component. In fact, at least 20% of human genes are regulated by miRNAs [Xie et al., 2005]. miRNAs are crucial in development, including zebrafish embryonic development [Wienholds et al., 2005], fruitfly's synthesis of protein crucial for memory formation [Ashraf et al., 2006], and plants' adaptive response to stresses [Mallory and Vaucheret, 2006], etc. miRNA also plays an important role in genetic diseases including cancer [Croce and Calin, 2005; Esquela-Kerscher and Slack, 2006].

To understand the biogenesis of miRNA is an important part of miRNA study. It is commonly believed that the primary transcript of miRNA gene (pri-miRNA) forms a stem-loop structure with two flanking ssRNA tails, which will then be cleaved so that pre-miRNA stem-loop structure is formed. The loop region will also be cleaved afterwards, resulting in miRNA:miRNA* duplex. Helicase will then unwind the duplex, and one of the strands, namely mature miRNA, will be incorporated into RISC complex to realize regulatory functions, while miRNA* is degraded. [Bartel, 2004; Lee et al., 2003; Denli et al., 2004; Kim and Kim, 2007]. Khvorova et al. and Schwarz et al. showed that the mature miRNA strand usually corresponds to the stand with less stable 5' ending [Khvorova et al., 2003; Schwarz et al., 2003], providing a theory of how the functional strand is selected. The less frequent event that both strands act as mature miRNAs could also happen. [Zeng et al., 2005] suggested that human Drosha cleaves approximately two helical RNA turns (~22bp) from the junction of pri-miRNA loop and the adjacent stem into the stem to produce pre-miRNA. Han et al. [Han et al., 2006; Seitz and Zamore, 2006] proposed an alternative model of human miRNA maturization, saying that cleavage site is determined as ~11bp from the stem-ssRNA tail junction.

Despite of all those studies, miRNA biogenesis even in human is still far from clear to us, let alone in other species. Whether are miRNA maturization pathways conserved across species is an example of question. So far people have already known that pre-miRNA is produced by Drosha cleavage in animals and by DCL1 in plants [Bartel, 2004]. Does DCL1 also rely on the ~11bp or ~22bp structural information to determine the cleavage site? It is a reasonable hypothesis that certain miRNAs are conserved and share similar maturization mechanisms in at least a few species, because many miRNA target genes are conserved across species [John et al., 2004]. The question is to what extent are they conserved.

On the other hand, to locate mature miRNA genes is important, for target regulation totally relies on mature miRNA but not its precursors (this article will adhere to the convention that precursor actually means extended pre-miRNA). Up to date, many algorithms are designed in complementary to biological experiments to predict miRNA genes from genome sequence [for example, Lim et al., 2003a, 2003b; Lai et al., 2003; Wang et al., 2005; Xue et al., 2005; Nam et al., 2005; Berezikov et al., 2005, 2006] and to predict miRNA targets given mature miRNA genes [for example, Rhoades et al., 2002; Enright et al., 2003; Stark et al., 2003; Wang et al., 2004; Rehmsmeier et al., 2004; Lewis et al., 2003, 2005]. However, very few researchers focused on predicting mature miRNA instead of its extended sequence (e.g. pri-miRNA). [Helvik et al., 2007] did the only work known to the author in this area, in which 33.4% of the 5' Drosha cleavage sites for 449 (11 pre-excluded) human miRNAs listed in miRBase 8.1 [Griffiths-Jones, 2004; Griffiths-Jones et al., 2006] are predicted completely accurate. Mature miRNA prediction not only reduces both labor and financial expenses of experiments, but also might be the only means to detect low-concentration miRNAs.

This article employs the fundamental idea of comparative genomics to attack these problems; it provides a first demonstration of the importance of the proposed thermodynamic and structural consensus principle. Interspecies conservation subgroups could be classified based on this principle. Meanwhile, a machine learning method is proposed to thermodynamically and structurally mimic the maturization of miRNA without using specific and detailed knowledge in human miRNA maturization (as used in [Helvik et al., 2007]), i.e., to provide the first example of generally predicting mature miRNA across species. Using 885 mature miRNA on 866 miRNA precursors as training samples, the algorithm predicts 79.4% of 2780 test mature miRNA genes on 2722 miRNA precursors in 44 species. For human miRNAs, the prediction rate is 84.7% on 346 test mature miRNA genes with the rest 127 mature miRNAs in miRBase 8.1 serving as training samples. (Helvik et al.'s human-only prediction method would produce a ~85% rate if the same scoring criterion is employed. See Results and Discussion.) Furthermore, this algorithm could be used to evaluate the degree of conservation of certain thermodynamic/structural features; higher prediction rate suggests deeper conservations. Then, mature miRNAs' location distribution on corresponding precursors is investigated using consensus structural information, and results suggest an important hypothesis on miRNA maturization mechanisms that largely deviates from current understanding [Zeng et al., 2005; Han et al., 2006]. miRNA structure prediction is also improved by using consensus thermodynamics; although consensus sequence and consensus structure are used in RNA structure prediction [Gardner and Giegerich, 2004], this article provides the first attempt of using consensus thermodynamics.

## Results and Discussion

*Conservation subgroups*

Given the total number of categories N, all miRNA precursors are assigned a category (labeled from {1,2,…N}) as well as a most probable conformation (using thermodynamic and structural consensus principle described in Materials and Methods). Although this subgroup categorization is done without introducing functional information, it reflects various biological and functional features of those miRNAs, showing that this categorization reasonably depicts conservation subgroups to certain level N.

Evolution information, for example, could be retrieved from the categorization. Each species is assigned the category (Table 1) which most miRNAs in this species belong to. When N=2, plants and animals are not separated, which is against the author's expectation. When N=7, insects and worms were grouped, mammals were divided into two groups, and plants were grouped but fused with one group of mammals. When N=10, insects and worms were grouped, plants were grouped, but primates were divided into several groups. Human was seldom grouped with the other primates, with one possible explanation that the number of human miRNAs overwhelms numbers of other primate miRNAs, making it look like a distinct category. The situation for N=44 is similar. Notice that if there wasn't any interspecies miRNAs conservation and miRNAs were only conserved within species, one should more or less expect each species was grouped into a distinct category, which is not the case. This justifies the existence of our discovery.

On the other hand, many miRNA families [Griffiths-Jones et al., 2005] are consistent with the subgroup categorization. For instance when N=7, 46 members of *mir-156* plant family [MIPF:MIPF0000008; Reinhart et al., 2002; Ambros, 2001] belong to subgroup 1, while only 6 of the rest 9 members belong to subgroup 3 and the other 3 belong to subgroup 5, 6, 7 respectively. Another example is *mir-2* family [MIPF:MIPF0000049; Ambros, 2001; Lee and Ambros, 2001], with 13 members in subgroup 6, 2 in subgroup 1, respectively in subgroup 2,3,5,7.

Moreover, target analysis agrees with the subgroup categorization. For instance, *SARA1*, COPII-associated small GTPase, is a target gene of *miR-106*, *miR-17–5p*, *miR-20* and miR-203 [John et al., 2004], while *miR-106*, *miR-17–5p* and *miR-20* are categorized in subgroup 5 and *miR-203* is the only exception when N=7; *SOS2*, son of sevenless protein homolog 2, is a target gene of *miR-98*, *let-7i,7e*, *miR-103*, *miR-107* and *miR-134* [John et al., 2004], while *miR-98*, *let-7i,7e*, *miR-103* and *miR-107* are categorized in subgroup 5 and *miR-134* is the only exception when N=7. It is reasonable to suppose miRNAs which regulate a same gene are evolutionarily close, and hence the above analysis provides another hint that the categorization reveals conservation subgroups.

It will be further seen that the subgroups are biologically meaningful.

*Predicting mature miRNA in 44 species*

Both thermodynamic and structural features are learned and then employed to select a tentative mature miRNA which is most similar to other members in corresponding conservation subgroup.

There are several conservation subgroup categorization schemes which could be used in predicting mature miRNA. Species could of course serve as categories. A coarser categorization will be the conservation subgroups learned from the data. The coarsest scheme will be to regard all 3588 miRNAs as one subgroup.

Prediction result when species are treated as categories is shown in Table 2. The prediction rate is 79.4% on 2780 test samples, 80.1% on all 3665 samples, and 82.6% on 885 test samples. As discussed in Materials and Methods, the prediction rate on test samples largely deviates from 100% and is not significantly higher than on testing samples. This might be because more than one maturization mechanisms exist even within one conservation subgroup. Prediction rate on 346 human testing samples is 84.7%, and is generally high for animal mature miRNAs, ranging from 80% to 95%. Prediction is much less accurate for plants miRNAs.

Prediction result when conservation subgroups learned from the data are regarded as categories is shown in Table 3. The overall prediction rates are between 65% and 70%, lower than when species are treated as categories. This is reasonable because species category depicts finer conservation subgroups. The prediction is meaningful despite of the low rate, since we want to study the inter-species conservation phenomena. Notice there is/are always one or two subgroup(s) on which prediction rate(s) is/are especially low. They/it represent(s) badly conserved miRNAs (at least in the thermodynamic and structural sense of conservation). For example, when N=7, worst predicted subgroup 1 is the group of plants miRNAs, whose abnormality will be further discussed later. Furthermore, as N increases, the prediction rate on the worst subgroup decreases while increases on the other subgroups. This is because the subgroup which consists of badly conserved miRNAs grows finer and finer. The prediction rates on many other subgroups reach 80% or more, as high as in the case where species act as categories. Since miRNA maturization and regulatory mechanisms should be relatively conserved within species [Lagos-Quintana et al., 2001; Mourelatos et al., 2002; Park et al., 2002; Reinhart et al., 2002; Lee and Ambros, 2001; Lau et al., 2001], a prediction rate close to the one when species acts as categories shows that those subgroups captured the inter-species conservation. If one excludes the badly conserved subgroup which mainly corresponds to plants, ~80% prediction rate could be obtained.

One may argue that since the subgroup categorization is learned while mature miRNA already known, the prediction based on this categorization will produce an inflated prediction rate.

Moreover, if one is given a new unlabeled miRNA, how could he/she determine which subgroup that miRNA belongs to. To clarify the first argument, please be aware that the purpose of predicting mature miRNAs using learned conservation subgroup categorization is not to demonstrate a high prediction rate but to show that this categorization indeed characterizes conservations groups. Indeed, prediction rate will be much lower if a random categorization is used and therefore the learned categorization makes sense. If a highly accurate prediction is needed, one could always use species as category. One discussion on the second argument would be: had one been given an unknown miRNA, he/she usually knows which species is the miRNA from, and therefore could find the corresponding subgroup. If not, to enumerate every possible category would be the strategy.

Mature miRNA could also be predicted without using category information, i.e. to regard all 3665 miRNAs as in one large conservation subgroup. In this case, prediction rates on testing set, training set and all data are 67.0%, 68.5% and 67.3% respectively. These are much lower values comparing to ~80% when conservation subgroups are used, since in the latter case subgroup categorization captured conservation subgroups information. It is worth mention that plant miRNAs are still predicted poorly.

The default setting under which the above results are produced is that a prediction will be regarded as wrong if either of the predicted 5' or 3' ending of mature miRNA was shifted more than $T=4$ nucleotides from the real locus. The prediction power will decrease significantly if $T$ becomes smaller (Figure. 1). Therefore, although the proposed method could accurately narrow down the location of mature miRNA to a small range, it performs badly if the exact mature miRNA 5' and 3' ending positions are asked. Hence none of the two existing methods for predicting mature miRNAs could be directly employed to predict targets. One strategy to bypass this obstacle could be to shift several nucleotides from the predicted mature miRNA to generate alternative mature miRNA candidates, and then use combinatorial approach to identify the target genes [Krek et al., 2005].

In order to predict the exact 5' and 3' endings of certain mature miRNA, more biological knowledge on specific maturization mechanism in corresponding conservation subgroup is needed. One example is [Helvik et al., 2007], in which mature miRNA prediction is the sole aim and therefore lots of detailed features in human miRNA maturization such as "nucleotide occurrences at each position in the 24 nucleotide regions of the precursor 5' and 3' arms" are employed. Their accuracy of exact prediction on human miRNAs is 33.4%, higher than the proposed algorithm's accuracy of 25.2%. It is not hard to believe that analogous features for various conservation subgroups could be studied, fine tuned and incorporated into our proposed method, and more sophisticated machine learning frameworks could be introduced(such as SVM used by Helvik et al.), resulting in a significant boost in the predicting power. Unfortunately, so far miRNAs maturization in many species other than human are not well studied, let alone in conservation subgroups. Therefore, separately studying detailed characteristic features of conservation subgroups will be one future direction, and higher prediction accuracy will be one of by-products.

*Consensus principle improves miRNA structure prediction*

Although there are many well-recognized RNA 2D-structure prediction algorithms (for example, mfold [Mathews et al., 1999; Zuker, 2003], RNAfold [Hofacker, 2003], and [Rivas and Eddy, 1999]), the predicted minimum free energy structure might not be the in vivo conformation. It is possible that after projecting 3D structure into a 2D plane the minimum energy structure no longer obtains lowest energy in the 2D conformation space. Another possibility is that in vivo RNA might be under certain interactions with the environment which might not only consist of water but enzymes and other molecules. In other words, Gibbs distribution for isolated system that solely depends on free energy might not hold here, and RNA in vivo might not obtain its lowest energy. In addition, parameters and model which are used to calculate energy of the molecule are only approximations. However, thanks to existing sophisticated 2D-structure prediction methods, it is most probable that the conformation in vivo corresponds to either the optimal structure or some predicted suboptimal structure whose energy is slightly higher than the optimal structure. Thermodynamic consensus within conserved miRNA subgroups is employed to pick suboptimal structure as the best approximation of in vivo structure.

If conformations of miRNAs used in the prediction are not determined by consensus principle but simply by minimum free energy, the prediction rate will be significantly reduced. For instance, when all 3665 mature miRNAs are considered as in one conservation subgroup, if conformations of miRNAs in the training set are chosen to be optimal structures but no longer the consensus structures, prediction rate on 2780 testing data will decrease from 67.0% to 59.9%. If in addition conformations of miRNAs in testing set are constrained to be only the minimum energy ones (suboptimal structures are no longer enumerated in the procedure of finding most consensus ones), the prediction rate will further decrease to 57.1%. Other schemes of conservation subgroup categorizations produce similar results.

Therefore, consensus principle really helps to improve prediction. Since the mature miRNA data is obtained by experiments, improved prediction implies that consensus principle helps to find in vivo conformations.

In fact, structural consensus is widely used in protein structure prediction (for example, see homology modeling [Schwede et al., 2003]). Researchers recently adopted similar approach for RNA structure prediction (reviewed in [Gardner and Giegerich, 2004]), using either sequence consensus or structural consensus or both. Here I provided the first example that thermodynamics consensus could also be one useful criterion. The computational complexity using thermodynamic consensus is between sequence consensus and structural consensus. Since the former is fast but not general while the latter is too slow, thermodynamic consensus is a possible compromise.

*Mature miRNA location*

Locations of all 3665 mature miRNA on corresponding precursors' stem-loop structures could be calculated using structures predicted by consensus principle. Basic statistical analysis is then employed to investigate features including mature miRNA's lengths, distances from mature miRNA to loop and to ssRNA tails. These statistics are summarized by species [Table 4]. In human, for instance, mature miRNAs are ~11 nucleotides away from ssRNA tails, although the standard deviation of this relative position is rather big (~7 nucleotides). The average distance between human mature miRNAs and corresponding center loops, on the other hand, is 5.67 with a fluctuation of 6.24 nucleotides. Taking into account that the fluctuation of the lengths of mature miRNAs is very small, the Drosha cleavage site is ~27 nucleotides from the center loop. This result slightly differs from [Zeng et al., 2005] but supports [Han et al., 2006].

However, this structural feature might not generally hold for other species. As shown in Table 4, either the distance between mature miRNA and ssRNA tails or the distance between mature miRNA and the center loop varies significantly across species. In fact, the ~11 nucleotide rule doesn't hold even within mammals. This result is not surprising because standard deviations are large too.

Although none of the two distances described above seems to be a conserved structural feature of miRNA maturization, the proportions of such distances to the length of the stem region in pri-miRNA are better characteristics. In order to save unnecessary writings, the distance between mature miRNA and ssRNA tails is denoted by D1 and the distance between mature miRNA and the center loop is denoted by D3, while D1 divided by stem length is denoted by D2, and similarly normalized D3 is denoted by D4. The sums of standard deviations of D1, D2, D3, D4 in all 44 species are respectively 268.08, 5.49, 350.72, 4.72. Although one should not directly compare the fluctuations of D1 and D3 with D2 and D4, other meaningful facts could be observed, namely 268.08<350.72 and 5.49>4.72. The former inequality suggests that mature-tail distance is more conserved than mature-loop distance, supporting [Han et al., 2006]'s argument over [Zeng et al., 2005]'s. The latter inequality, however, states the reverse direction.

In order to solve this paradox, the prediction algorithm is used to investigate the conservation degrees of these four structural features, since prediction power reflects conservation degrees of employed features, while a summation of standard deviations might convey biased information due to unbalanced species sizes. Using these four structural features respectively in the distance function as structural part of the consensus principle (please refer to Materials and Methods), four different predicting rates are obtained, namely 58.56%, 69.28%, 74.17%, 79.35% for D1, D2, D3, D4. One finds out that the distance between mature miRNA and the center loop is a more conserved structural feature, and the normalized value is more conserved than the absolute value. In this sense, [Zeng et al., 2005] might not be wrong as [Han et al., 2006] suggest. Furthermore, it puts forward an interesting hypothesis that certain mechanism which measures the proportional value but not the absolute distance controls the pre-miRNA biogenesis.

*Plant miRNAs are distinct from animal miRNAs*

A careful investigation tells that value of D3 is on average much larger in plant miRNAs than in animals, and it fluctuates more fiercely in plant. The prediction rate for plant miRNAs is generally much lower than for animal miRNAs too. It is already widely acknowledged that plant miRNAs are not conserved with animal miRNAs, with the conservation subgroup categorization reinforced this point. Furthermore, the large standard deviations of miRNAs within certain plant groups as well as the low prediction rate suggest that plant miRNAs might not even be conserved within species. Indeed, under the thermodynamic and structural metric described above, miRNAs are most sparsely distributed in the conservation subgroup which most plant miRNAs belong to. This coincides with the discovery that presence of miRNA clusters is very common in animals but not in plants [Millar and Waterhouse, 2005].

Furthermore, D3 and D4 are much larger for plant miRNAs than for animal miRNAs. The difference is so distinct that one naturally suspects that in animals and plants, different structural features are relied on in order to locate cleavage sites. This is a reasonable hypothesis since the pre-miRNA biogenesis involves Drosha in animals but DCL1 in plants. In order to investigate what structural features might be relied upon in animals and plants, their degrees of conservation are studied according to the predictive power of the described algorithm. Indeed, if D4 is used for animals and D2 is used for plants in the consensus principle, the highest prediction rate among all combinations could be obtained, namely 81.3% on 2780 test samples, higher than the value of 79.4% when D4 is used for both animals and plants. Considering the fact that plant miRNAs only constitute one quarter of the test data, this is already a significant boost and therefore tells us that D4 is more conserved for animal miRNAs while D2 is for plants. Since conservation stands for evolutionarily fixed feature, and housekeeping features are functional in most circumstances, if we stick to the existing belief that structural and thermodynamic features determine cleavage sites, a hypothesis on different mechanisms of miRNA maturization could be obtained. That is, Drosha recognizes cleavage site by loop-mature miRNA distance normalized by stem length and DCL1 recognizes cleavage site by tail-mature miRNA distance normalized by stem length.

## Materials and Methods

After excluding identical sequences, 3588 miRNA precursors with 3665 mature miRNA genes in 44 species are obtained from miRBase version 8.1. A random portion of the data is taken as training set, while the rest is taken as testing data. Although it is common in machine learning that the size of training set greatly affects the predictive power of the classifier, it is not the case in this specific problem. Unlisted data show that the prediction rates based on training sets respectively sampled with sampling probability 0.1, 0.2, and 0.3 are almost the same. Moreover, the algorithm is also tested on the training set, and the prediction rate is not approximately 100% but almost the same with on the testing set (please refer to Results and Discussion). Therefore it is possible to provide a good predictor with a small group of training data. As long as every conserved miRNA subgroup is sampled in the training set, the predictive power will not depend on the size much. In experiments stated in this article, training set is sampled with probability of 0.2, resulting in a training set of 866 miRNA precursors and 885 mature miRNA genes.

Optimal and several suboptimal structures of miRNA precursors in the data set are predicted using the well-recognized software mfold. Mfold outputs not only structures but also free energy value of every nucleotide for each structure, which will serve as the thermodynamics data.

For given conformations of two miRNA precursors, a distance between them is defined by a scaled summation of their free energy differences and the difference of their mature miRNA location on respective stem-loop structures, with a weight focusing the free energies of 5' 1-4, 11 sites and another weight balancing the thermodynamic part and structural part. Free energy difference is defined by a Euclidean distance between interpolated free energy vectors, whose element is the free energy value of each nucleotide in the mature miRNA region. The vectors are interpolated to same dimension since mature miRNAs' lengths might differ. Results only change slightly if distance definition is mildly modified but still including both thermodynamic and structural information (for example, change the Euclidean 2-norm to 1-norm). This suggests robustness of the proposed method, and results are not coincident or artificial.

Every miRNA precursor in the training set could be assigned to a category and a specific conformation chosen from its (sub)optimal structures. An assignment is searched so that distances within categories are minimized. By doing this, miRNAs are categorized into conservation subgroups and consensus conformations are picked at the same time. One could also fix category tags as species, in which case only consensus conformations are searched.

Then mature miRNAs' positions on their precursors are studied. Several indexed are investigated, including mature miRNA's distances to ssRNA tail, to loop, stem's length. Mean and standard deviation for each category are calculated.

We now predict mature miRNA location on its precursors. Focusing on biophysical properties crucial in biological pathways other than machine learning techniques, a simple K-neighborhood model is employed for learning and prediction. Given a precursor with mature miRNA location unknown, a window of 21bp is slided along the precursor's sequence, representing tentative mature miRNAs. Different conformations of the test miRNA precursor are enumerated, too. For each tentative mature miRNA with certain conformation, distances between it and mature miRNAs in the training set in the same conservation subgroup are calculated. K shortest distances are summed, representing the score of this enumeration. Tentative mature miRNA and corresponding precursor conformation with the smallest score is outputted as the prediction. This K-neighborhood classifier, though simple, captures the multi-category and nonlinear characteristic of the problem. Unlisted data shows that the value of K (1,2,3,5) will only affect the predictive power slightly, and therefore K is set to be 1 in stated experiments.

*Determining loop, tails and stems*

Given a structure, loop is defined as the unpaired region where pairings are most evenly distributed on its 5' and 3' flanking regions of 15 bp length.

5' tail is the longest possible region starting from the 5' ending of pri-miRNA where nucleotides are not paired with nucleotides in 3' downstream of the loop. 3'tail is similarly defined.

Stems are the part between loop and tails.

# Acknowledgements

MT did part of this work while he was in Tsinghua University, Beijing, China. He would like to thank Paul Steinberg, Michael Zhang, Michael Zuker, Titus Brown, Xuegong Zhang, Yanda Li, Xiaowo Wang and Chenghai Xue for valuable discussions.

Table 1.

| miRBase species id | Species name | N=2 | N=7 | N=10 | N=44 |
|---|---|---|---|---|---|
| cel | C. elegans | 1 | 6 | 4 | 18 |
| hsa | H. sapiens | 1 | 3 | 7 | 38 |
| dme | D. melanogaster | 2 | 6 | 4 | 18 |
| mmu | M. musculus | 1 | 1 | 3 | 18 |
| ath | A. thaliana | 1 | 1 | 3 | 35 |
| cbr | C. briggsae | 1 | 6 | 4 | 18 |
| rno | R. norvegicus | 1 | 1 | 3 | 4 |
| osa | O. sativa | 1 | 1 | 3 | 35 |
| ebv | Epstein Barr virus | 1 | 1 | 3 | 10 |
| gga | G. gallus | 1 | 1 | 4 | 42 |
| dps | D. pseudoobscura | 2 | 6 | 4 | 18 |
| dre | D. rerio | 1 | 6 | 4 | 18 |
| xla | Xenopus laevis | 2 | 6 | 4 | 18 |
| zma | Z. mays | 1 | 1 | 3 | 35 |
| sbi | Sorghum bicolor | 1 | 1 | 3 | 35 |
| oar | Ovis_aries | 1 | 2 | 6 | 9 |
| ame | A. mellifera | 1 | 6 | 4 | 18 |
| aga | A. gambiae | 1 | 6 | 4 | 18 |
| cfa | C. familiaris | 1 | 6 | 4 | 44 |
| mgh | Mouse gammaherpesvirus 68 | 1 | 1 | 2 | 35 |
| hcm | Human cytomegalovirus | 1 | 1 | 3 | 4 |
| mtr | Medicago truncatula | 1 | 1 | 3 | 21 |
| sof | Saccharum officinarum | 1 | 1 | 2 | 19 |
| gma | Glycine max | 1 | 1 | 3 | 21 |
| ptc | Populus trichocarpa | 1 | 1 | 3 | 35 |
| ssc | Sus scrofa | 1 | 3 | 3 | 21 |
| ksh | Kaposi sarcoma-associated herpesvirus | 1 | 6 | 2 | 18 |
| mml | M. mulata | 1 | 3 | 7 | 4 |
| ggo | Gorilla gorilla | 1 | 3 | 4 | 18 |
| ppy | Pongo pygmaeus | 1 | 5 | 2 | 18 |
| ppa | Pan_paniscus | 1 | 1 | 2 | 42 |
| age | Ateles geoffroyi | 1 | 5 | 2 | 4 |
| ptr | P. troglodytes | 1 | 1 | 4 | 18 |
| lla | Lagothrix lagotricha | 1 | 5 | 4 | 4 |
| mne | Macaca nemestrina | 1 | 1 | 3 | 4 |
| sla | Saguinus labiatus | 1 | 5 | 4 | 4 |
| lca | Lemur catta | 1 | 6 | 4 | 18 |
| fru | F. rubripes | 1 | 1 | 4 | 18 |

| tni | T. nigroviridis | 1 | 6 | 4 | 18 |
| --- | --- | --- | --- | --- | --- |
| ppt | Physcomitrella patens | 1 | 1 | 3 | 35 |
| sv4 | Simian virus | 1 | 5 | 2 | 4 |
| rlc | Rhesus lymphocryptovirus | 1 | 3 | 7 | 5 |
| bta | B. taurus | 1 | 3 | 3 | 38 |
| xtr | X. tropicalis | 1 | 6 | 6 | 18 |

Table 1. Categorization of species. Each of 44 species is assigned a category number from 1 to N, representing a conservation subgroup which that species belongs to. When N=7, for example, insects and worms were grouped, mammals were divided into two groups, and plants were grouped but fused with one group of mammals.

Table 2.

| | # Correct Prediction on Testing Data | # Testing Samples | Prediction Rate on Testing Data | # Correct Prediction on All Data | # All Samples | Prediction Rate on All Data |
|---|---|---|---|---|---|---|
| all: | 2206 | 2780 | **79.35%** | 2937 | 3665 | 80.14% |
| cel: | 71 | 84 | **84.52%** | 97 | 114 | 85.09% |
| hsa: | 293 | 346 | **84.68%** | 409 | 473 | 86.47% |
| dme: | 49 | 61 | **80.33%** | 65 | 79 | 82.28% |
| mmu: | 215 | 263 | **81.75%** | 288 | 348 | 82.76% |
| ath: | 47 | 91 | **51.65%** | 58 | 118 | 49.15% |
| cbr: | 48 | 58 | **82.76%** | 68 | 79 | 86.08% |
| rno: | 167 | 179 | **93.30%** | 220 | 240 | 91.67% |
| osa: | 62 | 134 | **46.27%** | 79 | 175 | 45.14% |
| ebv: | 22 | 22 | **100.00%** | 31 | 31 | 100.00% |
| gga: | 107 | 119 | **89.92%** | 132 | 146 | 90.41% |
| dps: | 50 | 55 | **90.91%** | 67 | 74 | 90.54% |
| dre: | 174 | 193 | **90.16%** | 243 | 272 | 89.34% |
| xla: | 0 | 6 | **0.00%** | 0 | 7 | 0.00% |
| zma: | 33 | 81 | **40.74%** | 43 | 97 | 44.33% |
| sbi: | 29 | 56 | **51.79%** | 37 | 72 | 51.39% |
| oar: | 2 | 2 | **100.00%** | 3 | 4 | 75.00% |
| ame: | 18 | 20 | **90.00%** | 23 | 25 | 92.00% |
| aga: | 26 | 28 | **92.86%** | 36 | 38 | 94.74% |
| cfa: | 3 | 4 | **75.00%** | 5 | 6 | 83.33% |
| mgh: | 0 | 9 | **0.00%** | 0 | 10 | 0.00% |
| hcm: | 6 | 10 | **60.00%** | 9 | 13 | 69.23% |
| mtr: | 9 | 11 | **81.82%** | 13 | 16 | 81.25% |
| sof: | 2 | 13 | **15.38%** | 2 | 16 | 12.50% |
| gma: | 11 | 18 | **61.11%** | 12 | 22 | 54.55% |
| ptc: | 93 | 159 | **58.49%** | 118 | 201 | 58.71% |
| ssc: | 35 | 40 | **87.50%** | 49 | 54 | 90.74% |
| ksh: | 11 | 12 | **91.67%** | 13 | 15 | 86.67% |
| mml: | 46 | 50 | **92.00%** | 55 | 61 | 90.16% |
| ggo: | 42 | 46 | **91.30%** | 57 | 62 | 91.94% |
| ppy: | 33 | 39 | **84.62%** | 47 | 54 | 87.04% |
| ppa: | 40 | 44 | **90.91%** | 48 | 53 | 90.57% |
| age: | 23 | 23 | **100.00%** | 33 | 33 | 100.00% |
| ptr: | 43 | 49 | **87.76%** | 57 | 63 | 90.48% |
| lla: | 21 | 24 | **87.50%** | 22 | 27 | 81.48% |
| mne: | 27 | 30 | **90.00%** | 37 | 41 | 90.24% |
| sla: | 15 | 16 | **93.75%** | 25 | 28 | 89.29% |
| lca: | 9 | 11 | **81.82%** | 14 | 16 | 87.50% |
| fru: | 88 | 103 | **85.44%** | 114 | 130 | 87.69% |
| tni: | 70 | 79 | **88.61%** | 94 | 104 | 90.38% |
| ppt: | 2 | 13 | **15.38%** | 4 | 18 | 22.22% |
| sv4: | 0 | 2 | **0.00%** | 0 | 2 | 0.00% |
| rlc: | 18 | 19 | **94.74%** | 21 | 22 | 95.45% |
| bta: | 25 | 26 | **96.15%** | 31 | 33 | 93.94% |
| xtr: | 121 | 132 | **91.67%** | 158 | 173 | 91.33% |

Table 2. Prediction rates when species is used as category. Rates for animal miRNAs roughly range from 80% to 95%, while rates for plants miRNAs are much lower. Notice prediction rate on testing data is similar to on all data (both training and testing), and in fact similar to on training data as well (not shown).

Table 3.

| N | 2 | 3 | 4 | 5 | 6 | 7 | 8 | 9 | 10 | 44 |
|---|---|---|---|---|---|---|---|---|---|---|
| Prediction rate on training set | 67.91% | 69.49% | 69.49% | 70.28% | 69.60% | 71.30% | 70.40% | 70.06% | 72.66% | 70.96% |
| Prediction rate on testing set | 66.19% | 66.62% | 66.73% | 66.33% | 66.55% | 67.23% | 67.16% | 68.60% | 68.60% | 69.42% |
| Overall prediction rate | 66.60% | 67.31% | 67.39% | 67.29% | 67.29% | 68.21% | 67.94% | 68.95% | 69.58% | 69.80% |
| Prediction rate on conservation subgroups of testing set: | | | | | | | | | | |
| 1 | **58.30%** | 80.17% | 85.47% | 79.47% | 81.25% | **39.50%** | 85.77% | **28.00%** | **28.40%** | - |
| 2 | | 83.68% | 85.63% | 84.97% | 85.34% | 73.73% | 87.90% | 89.80% | 90.79% | 81.58% | - |
| 3 | | | **49.44%** | **46.67%** | **41.47%** | 83.46% | 73.90% | 79.94% | 81.46% | **36.57%** | - |
| 4 | | | | 81.05% | 76.44% | 76.25% | 70.07% | 89.09% | **37.05%** | 90.87% | - |
| 5 | | | | | 83.46% | 85.00% | 81.23% | **35.48%** | 74.70% | 82.72% | - |
| 6 | | | | | | **40.97%** | 84.79% | **65.05%** | 80.67% | 81.02% | - |
| 7 | | | | | | | 76.83% | **38.40%** | 76.50% | 73.08% | - |
| 8 | | | | | | | | 71.86% | 70.32% | 70.94% | - |
| 9 | | | | | | | | | 87.93% | 72.64% | - |
| 10 | | | | | | | | | | 89.00% | - |

Table 3. Prediction rates when conservation subgroups learned from data are used as categories. Total subgroup number N ranges from 2 to 10 plus 44. Row 2-4 show prediction rates on 885 training data, 2780 testing data, and 3665 total data. Row 6-15 show prediction rates on each conservation subgroup of the testing data. N=44 case is not listed due to space limitation. Prediction with rate less than 70% is denoted by bold. Notice the sizes of conservation subgroups vary with different N values.

Table 4.

|   | E(D1) | Std(D1) | E(D2) | Std(D2) | E(D3) | Std(D3) | E(D4) | Std(D4) | E(Len) | Std(Len) |
|---|---|---|---|---|---|---|---|---|---|---|
| cel | 12.64 | 4.82 | 0.30 | 0.12 | 6.81 | 5.91 | 0.16 | 0.14 | 20.81 | 0.43 |
| hsa | 11.06 | 6.93 | 0.26 | 0.37 | 5.67 | 6.24 | 0.15 | 0.24 | 20.47 | 0.21 |
| dme | 9.37 | 4.86 | 0.24 | 0.11 | 6.10 | 5.13 | 0.15 | 0.10 | 21.53 | 0.53 |
| mmu | 9.80 | 6.83 | 0.25 | 0.14 | 5.60 | 5.81 | 0.14 | 0.14 | 20.56 | 0.25 |
| **ath** | **20.10** | **23.55** | **0.25** | **0.19** | **29.66** | **23.91** | **0.40** | **0.17** | **20.06** | **0.41** |
| cbr | 15.00 | 6.46 | 0.34 | 0.10 | 6.24 | 4.09 | 0.14 | 0.08 | 21.13 | 0.52 |
| rno | 11.51 | 5.19 | 0.29 | 0.11 | 4.87 | 3.01 | 0.13 | 0.08 | 20.79 | 0.30 |
| **osa** | **11.45** | **10.41** | **0.18** | **0.12** | **31.60** | **24.90** | **0.44** | **0.18** | **20.21** | **0.34** |
| ebv | 11.76 | 7.39 | 0.29 | 0.15 | 3.61 | 2.02 | 0.10 | 0.05 | 21.28 | 0.96 |
| gga | 10.23 | 5.60 | 0.27 | 0.12 | 4.39 | 3.33 | 0.12 | 0.09 | 20.96 | 0.38 |
| dps | 8.09 | 6.49 | 0.19 | 0.33 | 5.53 | 4.38 | 0.16 | 0.22 | 21.56 | 0.54 |
| dre | 14.28 | 10.01 | 0.32 | 0.15 | 4.63 | 4.25 | 0.11 | 0.11 | 21.03 | 0.28 |
| xla | 6.43 | 2.87 | 0.19 | 0.08 | 4.43 | 2.56 | 0.13 | 0.08 | 21.43 | 1.75 |
| **zma** | **13.39** | **6.71** | **0.21** | **0.11** | **33.26** | **33.66** | **0.45** | **0.15** | **19.69** | **0.45** |
| **sbi** | **11.78** | **8.75** | **0.19** | **0.14** | **27.71** | **17.07** | **0.44** | **0.15** | **19.79** | **0.52** |
| oar | 16.50 | 10.83 | 0.35 | 0.15 | 3.75 | 3.34 | 0.09 | 0.09 | 22.00 | 2.35 |
| ame | 12.12 | 3.76 | 0.31 | 0.07 | 3.56 | 3.23 | 0.09 | 0.08 | 21.92 | 0.94 |
| aga | 12.74 | 5.48 | 0.31 | 0.13 | 5.18 | 4.72 | 0.13 | 0.12 | 21.45 | 0.75 |
| cfa | 10.33 | 1.80 | 0.27 | 0.05 | 6.83 | 4.34 | 0.17 | 0.10 | 21.00 | 1.87 |
| mgh | -0.22 | 4.39 | -0.02 | 0.13 | 5.06 | 2.61 | 0.19 | 0.09 | 20.94 | 1.53 |
| hcm | 3.45 | 1.62 | 0.11 | 0.05 | 7.77 | 3.83 | 0.23 | 0.09 | 20.45 | 1.36 |
| **mtr** | **14.94** | **3.98** | **0.26** | **0.09** | **30.31** | **38.75** | **0.38** | **0.16** | **19.94** | **1.12** |
| **sof** | **16.25** | **1.56** | **0.17** | **0.08** | **74.06** | **36.46** | **0.61** | **0.18** | **19.88** | **1.11** |
| **gma** | **16.09** | **3.27** | **0.25** | **0.07** | **31.14** | **16.47** | **0.43** | **0.12** | **19.82** | **0.95** |
| **ptc** | **4.98** | **5.86** | **0.09** | **0.14** | **28.63** | **22.34** | **0.47** | **0.17** | **20.25** | **0.32** |
| ssc | 10.63 | 6.39 | 0.27 | 0.18 | 4.48 | 3.74 | 0.13 | 0.14 | 20.96 | 0.62 |
| ksh | 5.27 | 5.08 | 0.15 | 0.12 | 4.42 | 2.96 | 0.14 | 0.08 | 21.00 | 1.27 |
| mml | 11.69 | 6.37 | 0.29 | 0.12 | 4.69 | 2.87 | 0.12 | 0.08 | 20.86 | 0.59 |
| ggo | 13.23 | 6.98 | 0.31 | 0.12 | 4.52 | 2.75 | 0.11 | 0.07 | 20.87 | 0.58 |
| ppy | 12.74 | 6.78 | 0.31 | 0.13 | 4.60 | 2.64 | 0.12 | 0.07 | 20.77 | 0.63 |
| ppa | 13.50 | 6.56 | 0.32 | 0.11 | 4.52 | 3.01 | 0.11 | 0.08 | 20.81 | 0.63 |
| age | 11.69 | 5.81 | 0.30 | 0.10 | 3.78 | 3.04 | 0.10 | 0.09 | 21.13 | 0.81 |
| ptr | 12.98 | 6.75 | 0.32 | 0.12 | 4.22 | 2.59 | 0.11 | 0.07 | 20.97 | 0.59 |
| lla | 12.54 | 7.88 | 0.29 | 0.16 | 4.04 | 2.81 | 0.10 | 0.07 | 20.92 | 0.90 |
| mne | 12.48 | 6.26 | 0.30 | 0.11 | 4.63 | 2.96 | 0.12 | 0.07 | 20.98 | 0.72 |
| sla | 10.15 | 4.79 | 0.26 | 0.11 | 4.15 | 2.99 | 0.11 | 0.07 | 21.00 | 0.88 |
| lca | 9.93 | 4.52 | 0.27 | 0.08 | 3.53 | 2.80 | 0.09 | 0.08 | 21.00 | 1.18 |
| fru | 7.35 | 4.67 | 0.20 | 0.12 | 4.48 | 2.92 | 0.13 | 0.08 | 21.06 | 0.40 |
| tni | 6.84 | 5.80 | 0.18 | 0.16 | 4.48 | 3.06 | 0.13 | 0.09 | 21.06 | 0.45 |
| **ppt** | **12.94** | **11.23** | **0.18** | **0.16** | **36.28** | **18.57** | **0.49** | **0.17** | **20.11** | **1.06** |
| sv4 | 14.00 | 0.00 | 0.36 | 0.00 | 3.50 | 0.00 | 0.09 | 0.00 | 20.00 | 4.47 |
| rlc | 11.59 | 2.63 | 0.30 | 0.05 | 4.00 | 2.03 | 0.10 | 0.05 | 21.38 | 1.16 |
| bta | 8.39 | 4.75 | 0.23 | 0.12 | 3.91 | 2.95 | 0.11 | 0.08 | 21.00 | 0.80 |
| xtr | 9.59 | 5.41 | 0.25 | 0.12 | 4.16 | 3.67 | 0.12 | 0.10 | 20.81 | 0.35 |

Table 4. Mature miRNA locations on pri-miRNAs stem-loop structures. D1 stands for distance from mature miRNA to ssRNA tails, D2 is D1 divided by length of the stem which mature miRNA lies on, D3 stands for distance from mature miRNA to the center loop structure, D4 is D3 divided by length of the stem which mature miRNA lies on, Len is the length of mature miRNA. D1, D2, D3, D4, Len are all random variables, with 3665 data as samples. E(D1) and Std(D1) stand for D1's mean value and standard deviation. Bold identifies plants.

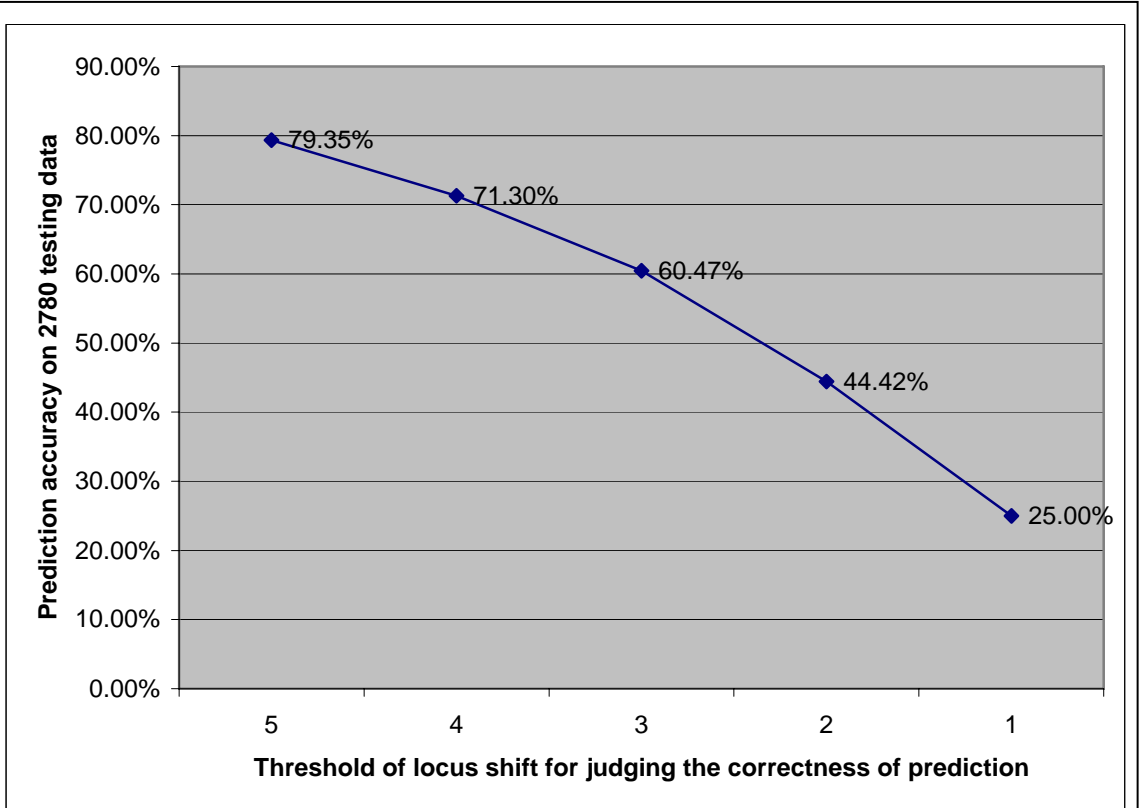

Figure 1. Prediction rate as a function of threshold T (locus shift). Prediction rate decreases significantly when more stringent criterion on the shift of predicted miRNA is applied.